\documentclass[12pt]{article}
\usepackage{graphicx}
\usepackage{lscape}
\usepackage{citesort}

\begin{document}

\def\ds{\displaystyle}
\def\beq{\begin{equation}}
\def\eeq{\end{equation}}
\def\bea{\begin{eqnarray}}
\def\eea{\end{eqnarray}}
\def\beeq{\begin{eqnarray}}
\def\eeeq{\end{eqnarray}}
\def\ve{\vert}
\def\vel{\left|}
\def\ver{\right|}
\def\nnb{\nonumber}
\def\ga{\left(}
\def\dr{\right)}
\def\aga{\left\{}
\def\adr{\right\}}
\def\lla{\left<}
\def\rra{\right>}
\def\rar{\rightarrow}
\def\nnb{\nonumber}
\def\la{\langle}
\def\ra{\rangle}
\def\ba{\begin{array}}
\def\ea{\end{array}}
\def\tr{\mbox{Tr}}
\def\ssp{{\Sigma^{*+}}}
\def\sso{{\Sigma^{*0}}}
\def\ssm{{\Sigma^{*-}}}
\def\xis0{{\Xi^{*0}}}
\def\xism{{\Xi^{*-}}}
\def\qs{\la \bar s s \ra}
\def\qu{\la \bar u u \ra}
\def\qd{\la \bar d d \ra}
\def\qq{\la \bar q q \ra}
\def\gGgG{\la g^2 G^2 \ra}
\def\q{\gamma_5 \not\!q}
\def\x{\gamma_5 \not\!x}
\def\g5{\gamma_5}
\def\sb{S_Q^{cf}}
\def\sd{S_d^{be}}
\def\su{S_u^{ad}}
\def\ss{S_s^{??}}
\def\sbp{{S}_Q^{'cf}}
\def\sdp{{S}_d^{'be}}
\def\sup{{S}_u^{'ad}}
\def\ssp{{S}_s^{'??}}
\def\sig{\sigma_{\mu \nu} \gamma_5 p^\mu q^\nu}
\def\fo{f_0(\frac{s_0}{M^2})}
\def\ffi{f_1(\frac{s_0}{M^2})}
\def\fii{f_2(\frac{s_0}{M^2})}
\def\O{{\cal O}}
\def\sl{{\Sigma^0 \Lambda}}
\def\es{\!\!\! &=& \!\!\!}
\def\ap{\!\!\! &\approx& \!\!\!}
\def\ar{&+& \!\!\!}
\def\ek{&-& \!\!\!}
\def\kek{\!\!\!&-& \!\!\!}
\def\cp{&\times& \!\!\!}
\def\se{\!\!\! &\simeq& \!\!\!}
\def\eqv{&\equiv& \!\!\!}
\def\kpm{&\pm& \!\!\!}
\def\kmp{&\mp& \!\!\!}


\def\simlt{\stackrel{<}{{}_\sim}}
\def\simgt{\stackrel{>}{{}_\sim}}


\title{
         {\Large
                 {\bf
Lepton polarization effects in $\Lambda_b \rar \Lambda \ell^+ \ell^-$ decay 
in family non--universal $Z^\prime$ model
                 }
         }
      }

\author{\vspace{1cm}\\
{\small T. M. Aliev \thanks {e-mail:
taliev@metu.edu.tr}~\footnote{permanent address:Institute of
Physics,Baku,Azerbaijan}\,\,, M. Savc{\i} \thanks
{e-mail: savci@metu.edu.tr}} \\
{\small Physics Department, Middle East Technical University,
06531 Ankara, Turkey }\\
       }

\date{}

\begin{titlepage}
\maketitle
\thispagestyle{empty}

\begin{abstract}
Possible manifestation of the family non--universal $Z^\prime$ boson
effects in lepton polarization in rare, exclusive baryonic $\Lambda_b \rar
\Lambda \ell^+ \ell^-$ decay is examined. It is observed that the double
lepton polarizations $P_{TT}$ and $P_{NN}$ are sensitive to the $Z^\prime$
contribution. Moreover, it is found that the zero position of the polarized
forward--backward asymmetry ${\cal A}_{FB}^{LL}$ is shifted to the left
compared to the standard model prediction. Therefore, determination of the
zero value of ${\cal A}_{FB}^{LL}$ is quite an efficient tool for
establishing new $Z^\prime$ boson, but also in discriminating various
scenarios of the considered family non--universal $Z^\prime$ model.
\end{abstract}

~~~PACS numbers: 13.30.--a, 14.20.Mr

\end{titlepage}

\section{Introduction}

Investigation of the rare decays described by the $b \to s(d)$ transitions
represents one of the main directions of high energy physics. The attractive
property of these decays is that they are forbidden at tree level in the
Standard Model (SM) and appear only at loop level. Therefore these decays
are quite promising for checking gauge structure of the theory at quantum
level. These decays are also excellent candidates in search of new physics
beyond the SM.

Rare decays in the $B$--meson sector described by $b \to s(d)$ transitions
have been studied theoretically (see for example \cite{R111201}and
references therein) and experimentally in detail (see for example
\cite{R111202}).

Exclusive $\Lambda_b \rar \Lambda \ell^+ \ell^-$, $\Lambda_b \rar \Lambda
\gamma$ decays in baryonic sector, which are described by $b \to s$
transition are also very interesting . The main advantage of these baryonic
decays is that, unlike mesonic decays, they can give information about the
helicity structure of the effective Hamiltonian \cite{R111203}. 

The baryonic decays $\Lambda_b \rar \Lambda \ell^+ \ell^-$, $\Lambda_b \rar
\Lambda \gamma$, $\Lambda_b \rar \Lambda \bar{\nu} \nu$ induced by the
flavor changing neutral current (FCNC) are studied comprehensively in many
works
\cite{R111204,R111205,R111206,R111207,R111208,R111209,R111202,R111210,R111211}.
The first step in experimental investigation of rare baryonic decays has recently
been taken by the CDF Collaboration, and they announced the observation of
the baryonic rare $\Lambda_b \rar \Lambda \mu^+ \mu^-$ decay. LHC--b
Collaboration is planning to study this decay in the near future \cite{R111213}.
The experimental observation of this decay has stimulated researches for a more refined
theoretical analysis of this subject. 

As has already been noted, rare decays induced by $b \to s$ transition are
quite promising for checking prediction of the SM and searching new physics
beyond the SM. In this sense, the physical observables like branching ratio,
forward--backward asymmetry ${\cal A}_{FB}$, single and double lepton
polarization effects, polarized forward--backward asymmetry are very useful.

Recently we have studied the rare $\Lambda_b \rar \Lambda \ell^+ \ell^-$ decay
within non--universal $Z^\prime$ model \cite{R111214}. The sensitivities of
the branching ratio, forward--backward asymmetry, and asymmetry parameters due to
the polarization of the $\Lambda$ and $\Lambda_b$ baryons, on $Z^\prime$
model parameters are investigated in detail.  

In the present work we perform an analysis of the single and double lepton
polarization effects, and polarized forward--backward asymmetries in the
framework of the non--universal $Z^\prime$ model developed in
\cite{R111215}. It should also be noted here that, so far, 
the effects of non--universal $Z^\prime$ model in the $B$--meson sector have 
been studied in many works \cite{R111216,R111217,R111218}.

The outline of the paper is as follows. In section 2 we present the
effective Hamiltonian responsible for the $b \to s \ell^+ \ell^-$
transition. In this section we also present the matrix element for the
$\Lambda_b \rar \Lambda \ell^+ \ell^-$ decay, and expressions of the
polarized forward--backward asymmetries in the $Z^\prime$ model. In section
3 the numerical results of these physical observables are given.   

\section{Theoretical framework}

Neglecting doubly Cabibbo--suppressed contribution, the effective
Hamiltonian responsible for the $b \to s \ell^+ \ell^-$ transition at
$\mu={\cal O}(m_b)$ scale is given as \cite{R111219} (see also the first
reference in \cite{R111201}),
\bea
\label{e111201}
H_{eff} = - {4 G_F \over \sqrt{2}} V_{tb}V_{ts}^\ast \sum_{i=1}^{10}
C_i(\mu) {\cal O}_i(\mu)~.
\eea

The expressions of the local operators ${\cal O}_i(\mu)$ can be found in
\cite{R111219} and the first reference in \cite{R111201}. The Wilson
coefficients are calculated in numerous works (see for example
\cite{R111220} and the references therein). The matrix element for the $b
\to s \ell^+ \ell^-$ transition in SM is given by,
\bea
\label{e111202}
M \es \frac{G_F \alpha_{em}}{2 \sqrt{2} \pi} V_{tb}V_{ts}^\ast \Bigg[
C_9^{eff}  \bar{s} \gamma_\mu (1-\gamma_5) b \, \bar{\ell}
\gamma_\mu \ell + C_{10} \bar{s} \gamma_\mu (1-\gamma_5) b \, \bar{\ell}
\gamma_\mu \gamma_5 \ell \nnb \\
\ek 2 m_b C_7 \bar{s} i \sigma_{\mu\nu} {q^\nu \over q^2} 
(1+\gamma_5) \, b \bar{\ell}
\gamma_\mu \ell \Bigg]~,
\eea
where $G_F$ is the Fermi constant, $\alpha_{em}$ is the fine structure
constant, $C_9^{eff}$, $C_{10}$ and $C_7$ are the relevant Wilson
coefficients. $V_{ij}$ are the elements of Kobayashi--Maskawa matrix.

The family non--universal $Z^\prime$ model considered in this work could
lead to FCNC at tree level, as well as to the appearance of new weak phases. 
Appearance of FCNS at tree level can be attributed to the non--diagonal
chiral coupling matrix. Assuming that the couplings of right--handed quarks
with $Z^\prime$ boson are flavor diagonal, and neglecting $Z$--$Z^\prime$
mixing, the $Z^\prime$ part of the effective Hamiltonian is given by,
\bea
\label{e111203}
H_{eff}^{Z^\prime} \es {2 G_F \over \sqrt{2}} V_{tb} V_{ts}^\ast \Bigg[
{B_{sb}^L B_{\ell\ell}^L \over V_{tb} V_{ts}^\ast } \bar{s} \gamma_\mu
(1-\gamma_5) b \, \bar{\ell} \gamma_\mu (1-\gamma_5) \ell \nnb \\
\ar {B_{sb}^L B_{\ell\ell}^R \over V_{tb} V_{ts}^\ast } \bar{s} \gamma_\mu
(1-\gamma_5) b \, \bar{\ell} \gamma_\mu (1+\gamma_5) \ell \bigg]~,
\eea
which can be rewritten as,
\bea
\label{e111204}
H_{eff}^{Z^\prime} = - {4 G_F \over \sqrt{2}} V_{tb} V_{ts}^\ast
\left( C_9^{Z^\prime} {\cal O}_9 + C_{10}^{Z^\prime} {\cal O}_{10} \right)~,
\eea
where 
\bea
\label{e111205}
C_9^{Z^\prime} \es - {g_S^2 \over e^2} {B_{sb}^L B_{\ell\ell}^R 
\over V_{tb} V_{ts}^\ast } S_{\ell\ell}^{LR}~, \nnb \\
C_{10}^{Z^\prime} \es {g_S^2 \over e^2} {B_{sb}^L
\over V_{tb} V_{ts}^\ast } {\cal D}_{\ell\ell}^{LR}~,
\eea
and,
\bea
\label{e111206}
S_{\ell\ell}^{LR} \es \left( B_{\ell\ell}^L + 
B_{\ell\ell}^R \right)~, \nnb \\
{\cal D}_{\ell\ell}^{LR} \es \left( B_{\ell\ell}^L - 
B_{\ell\ell}^R \right)~.
\eea
The off--diagonal element $B_{sb}^L$ might contain a new phase, and
therefore can be written as $\vel B_{sb}^L\ver  e^{i\varphi}$.

The essential point of this model is that $Z^\prime$ contribution does not
lead to the appearance of any new operators that exist in the SM, and its
contribution modifies the Wilson coefficients
$C_9$ and $C_{10}$. As a result, in order to take $Z^\prime$ effects into
account it is enough to make the following replacements in Eq.
(\ref{e111202}),
\bea
\label{e111207}
C_9^{eff} &\rar& C_9^{eff} - {4 \pi \over \alpha_S} (28.82) 
{B_{sb}^L \over V_{tb} V_{ts}^\ast } S_{\ell\ell}^{LR} = C_9^{tot}~, \nnb \\
C_{10} &\rar& C_{10} + {4 \pi \over \alpha_S} (28.82)
{B_{sb}^L \over V_{tb} V_{ts}^\ast } {\cal D}_{\ell\ell}^{LR}= C_{10}^{tot}~.
\eea

Our next task is to obtain the amplitude of the exclusive  $\Lambda_b \rar
\Lambda \ell^+ \ell^-$ decay. For this purpose we sandwich Eq.
(\ref{e111202}) between initial and final baryon states. Obviously, we need
to determine the matrix elements,
\bea
\label{nolabel}
&&\lla \Lambda(p) \vel \bar{s} \gamma_\mu (1-\gamma_5) b \ver \Lambda_b(p_B)
\rra~,~\mbox{\rm and,} \nnb \\
&&\lla \Lambda(p) \vel \bar{s} i\sigma_{\mu\nu} q^\nu (1+\gamma_5) b \ver 
\Lambda_b(p_B) \rra~. \nnb
\eea
These matrix elements are parametrized in terms of the form factors as
follows,
\bea
\label{e111208}
\lefteqn{
\lla \Lambda(p)\vel \bar{s} \gamma_\mu (1 - \gamma_5) b \ver \Lambda_b(p_B) 
\rra = \bar{u}_\Lambda (p) \Big[ f_1(q^2) \gamma_\mu + i f_2 (q^2)
\sigma_{\mu\nu} q^\nu } \nnb \\ 
\ar f_3(q^2) q_\mu - g_1 (q^2) \gamma_\mu\gamma_5 - i
g_2 (q^2) \sigma_{\mu\nu}\gamma_5 q^\nu - g_3 (q^2) \gamma_5 q_\mu 
\Big] u_{\Lambda_b} (p_B)~,\\ \nnb \\
\label{e111209}
\lefteqn{
\lla \Lambda(p) \vel \bar{s} i\sigma_{\mu\nu} q^\nu (1+\gamma_5) b
\ver\Lambda_b(p_B) \rra = \bar{u}_\Lambda (p) \Big[ f_1^T(q^2) \gamma_\mu +
i f_2^T (q^2) \sigma_{\mu\nu} q^\nu } \nnb \\ 
\ar f_3^T(q^2) q_\mu
+ g_1^T (q^2) \gamma_\mu\gamma_5 + i 
g_2^T (q^2) \sigma_{\mu\nu}\gamma_5 q^\nu + g_3^T (q^2) \gamma_5 q_\mu
\Big]u_{\Lambda_b} (p_B)~,
\eea
where $q^2=(p_B-p_\Lambda)^2$ and $f_i$, $g_i$, $f_i^T$, $g_i^T$ are 
the form factors responsible for the $\Lambda_b \to \Lambda$ transition.

Using Eqs. (\ref{e111207})--(\ref{e111209}), one can easily obtain the
matrix element of the $\Lambda_b \rar \Lambda \ell^+ \ell^-$ decay which is
given by,
\bea
\label{e111210}
M \es {G_F \alpha_{em} \over 4 \sqrt{2} \pi} V_{tb} V_{ts}^\ast \Bigg\{
\bar{\ell} \gamma_\mu \ell \, \bar{u}_\Lambda (p)
\Big[ A_1 \gamma_\mu (1+\gamma_5) + B_1 \gamma_\mu (1-\gamma_5)
+ i \sigma_{\mu\nu} q^\nu \Big( A_2 (1+\gamma_5) \nnb \\
\ar B_2 (1-\gamma_5) \Big)
+ q_\mu \Big(  A_3 (1+\gamma_5) + B_3 (1-\gamma_5) \Big) \Big] 
u_{\Lambda_b} (p_B) \nnb \\
\ar \bar{\ell} \gamma_\mu \gamma_5 \ell \, \bar{u}_\Lambda (p)
\Big[ D_1 \gamma_\mu (1+\gamma_5) + E_1 \gamma_\mu (1-\gamma_5) 
+ i \sigma_{\mu\nu} q^\nu \Big( D_2 (1+\gamma_5) + 
E_2 (1-\gamma_5) \Big) \nnb \\
\ar q_\mu \Big(  D_3 (1+\gamma_5) + E_3 (1-\gamma_5) \Big) \Big]
u_{\Lambda_b} (p_B) \Bigg\}~,
\eea
where
\bea
\label{nolabel}
A_1 \es - {2 m_b \over q^2} C_7 (f_1^T + g_1^T) + C_9^{tot} (f_1 - g_1)~, \nnb \\
A_2 \es A_1(1 \to 2)~,~~ A_3 = A_1(1 \to 3)~,\nnb \\
B_i \es A_i(g_i \to - g_i, g_i^T \to - g_i^T)~, \nnb \\
D_i \es C_{10}^{tot} (f_1 - g_1)~,~~D_2 \to D_1 (1 \to 2)~,~~D_3 \to D_1 
(1 \to 3)~ \nnb \\
E_i \es D_i (g_i \to - g_i)~. \nnb
\eea

The matrix element for the $\Lambda_b \rar \Lambda \ell^+ \ell^-$ decay
given in Eq. (\ref{e111210}) is the starting for us for all further
discussion. In order to calculate the double lepton polarization effects, we
introduce the orthogonal unit vectors $s_i^{\pm \mu}$ in the rest frame of
leptons,
\bea
\label{e111211}
s_L^{-\mu} \es \left(0, \vec{e}_L^{\,-} \right) = \left(0, {\vec{p}_- \over \vel
\vec{p}_- \ver } \right)~, \nnb \\
s_N^{-\mu} \es \left(0, \vec{e}_N^{\,-} \right) = \left(0, {\vec{p}_\Lambda
\times \vec{p}_- \over 
\vel \vec{p}_\Lambda \times \vec{p}_- \ver } \right)~, \nnb \\
s_T^{-\mu} \es \left(0, \vec{e}_T^{\,-} \right) = \left(0, \vec{e}_N \times
\vec{e}_L \right)~. 
\eea
The unit vectors for the polarizations of $\ell^+$ lepton can be obtained
from Eq. (\ref{e111211}) by making the replacement $\vec{p}_- \to
\vec{p}_+$.
Here, $\vec{p}_-(\vec{p}_+)$ and $\vec{p}_\Lambda$ are the three momenta of
the $\ell^-(\ell^+)$ lepton and $\Lambda$ baryon in the center of mass frame
(CM) of the lepton pair. Transformation of the unit vector $s_i^{\pm \mu}$
from rest frame to CM of the leptons can done by Lorentz boosting. It should
be noted here that, in performing Lorentz boosts transversal and normal
components are unchanged, and only longitudinal component $s_L^{\pm \mu}$ is
transformed. As a result we get,
\bea
\label{e111212}
\left( s_L^{\pm \mu} \right)_{CM} = \left( {\vel \vec{p}_\pm \ver \over
m_\ell}, {E_\ell \, \vec{p}_\pm \over m_\ell \vel \vec{p}_\pm \ver } \right)~.
\eea

Now we are ready to define the double lepton polarizations. Following
\cite{R111221} we define double and single lepton polarizations in the
following way,
\bea
\label{e111213}
P_{ij}(q^2) \es
\frac{ 
\Big( \ds \frac{d\Gamma(\vec{s}^{\,\,-}_i,\vec{s}^{\,\,+}_j)}{d \hat{s}} -
      \ds \frac{d\Gamma(-\vec{s}^{\,\,-}_i,\vec{s}^{\,\,+}_j)}{d \hat{s}} \Big) -
\Big( \ds \frac{d\Gamma(\vec{s}^{\,\,-}_i,-\vec{s}^{\,\,+}_j)}{d \hat{s}} -      
      \ds \frac{d\Gamma(-\vec{s}^{\,\,-}_i,-\vec{s}^{\,\,+}_j)}{d \hat{s}} \Big) 
     }
     {    
\Big( \ds \frac{d\Gamma(\vec{s}^{\,\,-}_i,\vec{s}^{\,\,+}_j)}{d \hat{s}} +      
      \ds \frac{d\Gamma(-\vec{s}^{\,\,-}_i,\vec{s}^{\,\,+}_j)}{d \hat{s}} \Big) +
\Big( \ds \frac{d\Gamma(\vec{s}^{\,\,-}_i,-\vec{s}^{\,\,+}_j)}{d \hat{s}} +      
      \ds \frac{d\Gamma(-\vec{s}^{\,\,-}_i,-\vec{s}^{\,\,+}_j)}{d \hat{s}} \Big)
     }~, \nnb \\ \nnb \\
P_{i}(q^2) \es
\frac{ 
      \ds \frac{d\Gamma(\vec{s}_i)}{d \hat{s}} - 
      \ds \frac{d\Gamma(-\vec{s}_i)}{d \hat{s}}
     }
     {
      \ds \frac{d\Gamma(\vec{s}_i)}{d \hat{s}} +
      \ds \frac{d\Gamma(-\vec{s}_i)}{d \hat{s}}
     }~.
\eea
The first (second) subindex of $P_{ij}$ represents polarization of
lepton (anti--lepton).

In this work we also investigate the polarized forward--backward
asymmetries, which are defined as,

\bea
\label{e111219}
{\cal A}_{FB}^{ij}(\hat{s}) \es 
\Bigg(\frac{d\Gamma(\hat{s})}{d\hat{s}} \Bigg)^{-1}
\Bigg\{ \int_0^1 dz - \int_{-1}^0 dz \Bigg\}
\Bigg\{ 
\Bigg[
\frac{d^2\Gamma(\hat{s},\vec{s}^{\,\,-}_i,\vec{s}^{\,\,+}_j)}
{d\hat{s} dz} - 
\frac{d^2\Gamma(\hat{s},\vec{s}^{\,\,-}_i,-\vec{s}^{\,\,+}_j)} 
{d\hat{s} dz}
\Bigg] \nnb \\
\ek
\Bigg[
\frac{d^2\Gamma(\hat{s},-\vec{s}^{\,\,-}_i,\vec{s}^{\,\,+}_j)} 
{d\hat{s} dz} - 
\frac{d^2\Gamma(\hat{s},-\vec{s}^{\,\,-}_i,-\vec{s}^{\,\,+}_j)} 
{d\hat{s} dz}
\Bigg]
\Bigg\}~,\nnb \\ \nnb \\
\es 
{\cal A}_{FB}(\vec{s}^{\,\,-}_i,\vec{s}^{\,\,+}_j)   -
{\cal A}_{FB}(\vec{s}^{\,\,-}_i,-\vec{s}^{\,\,+}_j)  - 
{\cal A}_{FB}(-\vec{s}^{\,\,-}_i,\vec{s}^{\,\,+}_j)  \nnb \\
\ar   
{\cal A}_{FB}(-\vec{s}^{\,\,-}_i,-\vec{s}^{\,\,+}_j)~.
\eea

Using the same convention and notations used in \cite{R111207},
for the double lepton polarizations we get,

\bea
\label{e111214}
P_{LL} \es \frac{16 m_{\Lambda_b}^4}{3 \Delta}
\mbox{\rm Re} \Bigg\{
- 6 m_{\Lambda_b} \sqrt{\hat{r}_\Lambda}
(1-\hat{r}_\Lambda+\hat{s})
\Big[ \hat{s} (1+v^2) (A_1 A_2^\ast + B_1 B_2^\ast)  - 
4 \hat{m}_\ell^2 (D_1 D_3^\ast + E_1 E_3^\ast) \Big] \nnb \\
\ar 6 m_{\Lambda_b} (1-\hat{r}_\Lambda-\hat{s})
\Big[ \hat{s} (1+v^2) (A_1 B_2^\ast + A_2 B_1^\ast) + 
4 \hat{m}_\ell^2 (D_1 E_3^\ast + D_3 E_1^\ast) \Big] \nnb \\
\ar 12 \sqrt{\hat{r}_\Lambda} \hat{s} (1+v^2) 
\Big( A_1 B_1^\ast + D_1 E_1^\ast +
m_{\Lambda_b}^2 \hat{s} A_2 B_2^\ast \Big) \nnb \\
\ar 12 m_{\Lambda_b}^2 \hat{m}_\ell^2 \hat{s} (1+\hat{r}_\Lambda-\hat{s})
\ga \vel D_3 \ver^2 + \vel E_3 \ver^2 \dr \nnb \\
\ek (1+v^2)
\Big[ 1+\hat{r}_\Lambda^2 - 
\hat{r}_\Lambda (2-\hat{s}) +\hat{s} (1-2 \hat{s}) \Big]
\Big(\vel A_1 \ver^2 + \vel B_1 \ver^2 \Big) \nnb \\
\ek \Big[
(5 v^2 - 3) (1-\hat{r}_\Lambda)^2 +   
4 \hat{m}_\ell^2 (1+\hat{r}_\Lambda) +
2 \hat{s} (1+8 \hat{m}_\ell^2 + \hat{r}_\Lambda)
- 4 \hat{s}^2 \Big] \Big( \vel D_1 \ver^2 + \vel E_1 \ver^2 \Big) \nnb \\
\ek m_{\Lambda_b}^2 (1+v^2) \hat{s}
\Big[2 + 2 \hat{r}_\Lambda^2 -\hat{s}(1 +\hat{s}) -
\hat{r}_\Lambda (4 + \hat{s})\Big] \Big(
\vel A_2 \ver^2 + \vel B_2 \ver^2 \Big) \nnb \\
\ek 2 m_{\Lambda_b}^2 \hat{s} v^2 \Big[
2 (1 + \hat{r}_\Lambda^2) - \hat{s} (1+\hat{s}) - 
\hat{r}_\Lambda (4+\hat{s})\Big] \Big(
\vel D_2 \ver^2 + \vel E_2 \ver^2 \Big) \nnb \\
\ar 12 m_{\Lambda_b} \hat{s} (1-\hat{r}_\Lambda-\hat{s}) v^2
\Big( D_1 E_2^\ast + D_2 E_1^\ast \Big) \nnb \\
\ek 12 m_{\Lambda_b} \sqrt{\hat{r}_\Lambda} \hat{s}
(1-\hat{r}_\Lambda+\hat{s}) v^2
\Big( D_1 D_2^\ast + E_1 E_2^\ast \Big) \nnb \\
\ar 24 m_{\Lambda_b}^2 \sqrt{\hat{r}_\Lambda} \hat{s}
\Big( \hat{s} v^2 D_2 E_2^\ast + 
2 \hat{m}_\ell^2 D_3 E_3^\ast \Big)
\Bigg\}~, \\ \nnb \\
\label{e111215}
P_{LN} \es - P_{NL} = \frac{16 \pi m_{\Lambda_b}^4 \hat{m}_\ell 
\sqrt{\lambda}}{\Delta \sqrt{\hat{s}}}  \mbox{\rm Im} \Bigg\{
(1-\hat{r}_\Lambda) 
(A_1^\ast D_1 + B_1^\ast E_1) \nnb \\
\ar m_{\Lambda_b}
 \hat{s} (A_1^\ast E_3 - A_2^\ast E_1 + B_1^\ast D_3
-B_2^\ast D_1) \nnb \\
\ar m_{\Lambda_b}
 \sqrt{\hat{r}_\Lambda} \hat{s}
(A_1^\ast D_3 + A_2^\ast D_1 +B_1^\ast E_3 + B_2^\ast E_1)
- m_{\Lambda_b}^2 \hat{s}^2
(B_2^\ast E_3 + A_2^\ast D_3)
\Bigg\}~, \\ \nnb \\
\label{e111217}
P_{LT} \es \frac{16 \pi m_{\Lambda_b}^4 \hat{m}_\ell \sqrt{\lambda} v}{\Delta \sqrt{\hat{s}}} 
\mbox{\rm Re} \Bigg\{
(1-\hat{r}_\Lambda) \Big( \vel D_1 \ver^2 + \vel E_1 \ver^2 \Big)
- \hat{s} \Big(A_1 D_1^\ast - B_1 E_1^\ast 
\Big) \nnb \\
\ek m_{\Lambda_b}
\hat{s}
\Big[ B_1 D_2^\ast + (A_2 + D_2 -D_3) E_1^\ast
-  A_1 E_2^\ast
-(B_2-E_2+E_3) D_1^\ast \Big]
\nnb \\
\ar m_{\Lambda_b}^2
\hat{s} (1-\hat{r}_\Lambda) 
(A_2 D_2^\ast - B_2 E_2^\ast) \nnb \\
\ar m_{\Lambda_b}
 \sqrt{\hat{r}_\Lambda} \hat{s}
\Big[ A_1 D_2^\ast + (A_2 + D_2 +D_3) D_1^\ast - B_1 E_2^\ast -
(B_2 - E_2 - E_3) E_1^\ast \Big] \nnb \\ 
\ek m_{\Lambda_b}^2 \hat{s}^2
(D_2 D_3^\ast + E_2 E_3^\ast )
\Bigg\}~, \\ \nnb \\
\label{e111218}
P_{TL} \es \frac{16 \pi m_{\Lambda_b}^4 \hat{m}_\ell \sqrt{\lambda} v}{\Delta \sqrt{\hat{s}}} 
\mbox{\rm Re} \Bigg\{
(1-\hat{r}_\Lambda) \Big( \vel D_1 \ver^2 + \vel E_1 \ver^2\Big)
+ \hat{s} \Big(A_1 D_1^\ast - B_1 E_1^\ast
\Big) \nnb \\
\ar m_{\Lambda_b}
\hat{s}
\Big[ B_1 D_2^\ast + (A_2 - D_2 + D_3) E_1^\ast
-  A_1 E_2^\ast
- (B_2+E_2-E_3) D_1^\ast \Big]
 \nnb \\
\ek m_{\Lambda_b}^2
\hat{s} (1-\hat{r}_\Lambda)
(A_2 D_2^\ast - B_2 E_2^\ast)
\nnb \\
\ek m_{\Lambda_b}
 \sqrt{\hat{r}_\Lambda} \hat{s}
\Big[ A_1 D_2^\ast + (A_2 - D_2 - D_3) D_1^\ast - B_1 E_2^\ast -
(B_2 + E_2 + E_3) E_1^\ast \Big] \nnb \\
\ek m_{\Lambda_b}^2
 \hat{s}^2
(D_2 D_3^\ast + E_2 E_3^\ast )
\Bigg\}~, \\ \nnb \\
\label{e111219}
P_{NT} \es - P_{TN} = \frac{64 m_{\Lambda_b}^4 \lambda v}{3 \Delta}
\mbox{\rm Im} \Bigg\{
(A_1 D_1^\ast +B_1 E_1^\ast)
+ m_{\Lambda_b}^2\hat{s}
(A_2^\ast D_2 + B_2^\ast E_2)
\Bigg\}~, \\ \nnb \\
\label{e111221}
P_{NN} \es \frac{32 m_{\Lambda_b}^4}{3 \hat{s} \Delta}
\mbox{\rm Re} \Bigg\{
24 \hat{m}_\ell^2 \sqrt{\hat{r}_\Lambda} \hat{s}
( A_1 B_1^\ast + D_1 E_1^\ast )
- 12 m_{\Lambda_b} \hat{m}_\ell^2 \sqrt{\hat{r}_\Lambda} \hat{s}  
(1-\hat{r}_\Lambda +\hat{s}) (A_1 A_2^\ast + B_1 B_2^\ast) \nnb \\
\ar 6 m_{\Lambda_b} \hat{m}_\ell^2 \hat{s} \Big[ 
m_{\Lambda_b} \hat{s} (1+\hat{r}_\Lambda-\hat{s})
\Big(\vel D_3 \ver^2 + \vel E_3 \ver^2 \Big) +
2 \sqrt{\hat{r}_\Lambda} (1-\hat{r}_\Lambda+\hat{s})
(D_1 D_3^\ast + E_1 E_3^\ast)\Big] \nnb \\        
\ar 12 m_{\Lambda_b} \hat{m}_\ell^2 \hat{s} (1-\hat{r}_\Lambda-\hat{s})
(A_1 B_2^\ast + A_2 B_1 ^\ast + D_1 E_3^\ast + D_3 E_1^\ast) \nnb \\
\ek [ \lambda \hat{s} + 
2 \hat{m}_\ell^2 (1 + \hat{r}_\Lambda^2 - 2 \hat{r}_\Lambda + 
\hat{r}_\Lambda \hat{s} + \hat{s} - 2 \hat{s}^2) ]
\Big( \vel A_1 \ver^2 + \vel B_1 \ver^2 - \vel D_1 \ver^2 - 
\vel E_1 \ver^2 \Big) \nnb \\
\ar 24 m_{\Lambda_b}^2 \hat{m}_\ell^2 \sqrt{\hat{r}_\Lambda} \hat{s}^2 
(A_2 B_2^\ast + D_3 E_3^\ast)
- m_{\Lambda_b}^2 \lambda \hat{s}^2 v^2 
\Big( \vel D_2 \ver^2 + \vel E_2 \ver^2 \Big) \nnb \\
\ar m_{\Lambda_b}^2 \hat{s} \{ \lambda \hat{s} -
2 \hat{m}_\ell^2 [2 (1+ \hat{r}_\Lambda^2) - \hat{s} (1+\hat{s})
- \hat{r}_\Lambda (4+\hat{s})]\}
\Big( \vel A_2 \ver^2 + \vel B_2 \ver^2 \Big)
\Bigg\}~, \\ \nnb \\
\label{e111222}
P_{TT} \es \frac{32 m_{\Lambda_b}^4}{3 \hat{s} \Delta}
\mbox{\rm Re} \Bigg\{
- 24 \hat{m}_\ell^2 \sqrt{\hat{r}_\Lambda} \hat{s}
( A_1 B_1^\ast + D_1 E_1^\ast )
- 12 m_{\Lambda_b} \hat{m}_\ell^2 \sqrt{\hat{r}_\Lambda} \hat{s}  
(1-\hat{r}_\Lambda +\hat{s}) (D_1 D_3^\ast + E_1 E_3^\ast) \nnb \\
\ek 24 m_{\Lambda_b}^2 \hat{m}_\ell^2 \sqrt{\hat{r}_\Lambda} \hat{s}^2
( A_2 B_2^\ast + D_3 E_3^\ast ) \nnb \\
\ek 6 m_{\Lambda_b} \hat{m}_\ell^2 \hat{s} \Big[ 
m_{\Lambda_b} \hat{s} (1+\hat{r}_\Lambda-\hat{s})
\Big(\vel D_3 \ver^2 + \vel E_3 \ver^2 \Big) -
2 \sqrt{\hat{r}_\Lambda} (1-\hat{r}_\Lambda+\hat{s})
(A_1 A_2^\ast + B_1 B_2^\ast)\Big] \nnb \\
\ek 12 m_{\Lambda_b} \hat{m}_\ell^2 \hat{s} (1-\hat{r}_\Lambda-\hat{s})
(A_1 B_2^\ast + A_2 B_1 ^\ast + D_1 E_3^\ast + D_3 E_1^\ast) \nnb \\
\ek [ \lambda \hat{s} - 
2 \hat{m}_\ell^2 (1 + \hat{r}_\Lambda^2 - 2 \hat{r}_\Lambda + 
\hat{r}_\Lambda \hat{s} + \hat{s} - 2 \hat{s}^2) ]
\Big( \vel A_1 \ver^2 + \vel B_1 \ver^2 \Big) \nnb \\
\ar m_{\Lambda_b}^2 \hat{s} \{ \lambda \hat{s} +
\hat{m}_\ell^2 [4 (1- \hat{r}_\Lambda)^2 - 2 \hat{s} (1+\hat{r}_\Lambda)         
- 2 \hat{s}^2 ]\}
\Big( \vel A_2 \ver^2 + \vel B_2 \ver^2 \Big) \nnb \\
\ar \{ \lambda \hat{s} -
2 \hat{m}_\ell^2 [5 (1- \hat{r}_\Lambda)^2 - 7 \hat{s} (1+\hat{r}_\Lambda)         
+ 2 \hat{s}^2 ]\}                                                               
\Big( \vel D_1 \ver^2 + \vel E_1 \ver^2 \Big) \nnb \\
\ek m_{\Lambda_b}^2 \lambda \hat{s}^2 v^2 
\Big( \vel D_2 \ver^2 + \vel E_2 \ver^2 \Big) 
\Bigg\}~.
\eea

Using the definition of single lepton polarization we find,

\bea
\label{e111223}
P_L^\mp \es {64 m_{\Lambda_b}^4 \hat{s} v \over \Delta} \Bigg\{
\pm  \sqrt{\hat{r}_\Lambda} \Big(
2 \, \mbox{\rm Re}[A_1^\ast E_1 + B_1^\ast D_1] -
m_{\Lambda_b} (1-\hat{r}_\Lambda+\hat{s}) \, \mbox{\rm Re}[A_1^\ast D_2+A_2^\ast D_1]                
\Big) \nnb \\
\kmp m_{\Lambda_b} \sqrt{\hat{r}_\Lambda} (1-\hat{r}_\Lambda+\hat{s}) \,
\mbox{\rm Re}[B_1^\ast E_2 + B_2^\ast E_1] \pm 2 m_{\Lambda_b}^2 \hat{s} \sqrt{\hat{r}_\Lambda} \,
\mbox{\rm Re}[A_2^\ast E_2 + B_2^\ast D_2] \nnb \\
\kpm m_{\Lambda_b} (1-\hat{r}_\Lambda-\hat{s}) \,\mbox{\rm Re}[A_1^\ast E_2 +
A_2^\ast E_1 + B_1^\ast D_2 + B_2^\ast D_1] \nnb \\
\kmp \frac{1}{3 \hat{s}} [ 1+\hat{r}_\Lambda^2 +\hat{r}_\Lambda (\hat{s}-2)+\hat{s}(1-2 \hat{s})] \,
\mbox{\rm Re}[A_1^\ast D_1 + B_1^\ast E_1] \nnb \\
\kmp \frac{1}{3} m_{\Lambda_b}^2 [2 + \hat{r}_\Lambda (2 \hat{r}_\Lambda -4 -\hat{s}) -\hat{s} (1+\hat{s})] \,
\mbox{\rm Re}[A_2^\ast D_2 + B_2^\ast E_2]
\Bigg\} ~, \nnb \\ \nnb \\
\label{e111224}
P_T^\mp \es {16 \pi m_{\Lambda_b}^3 \hat{m}_\ell \sqrt{\hat{s}\lambda} \over \Delta} 
\Bigg\{- \Big( \vel A_1 \ver^2  - \vel B_1 \ver^2 \Big)
+ 2 m_{\Lambda_b} \, \mbox{\rm Re}[A_1^\ast
B_2-A_2^\ast B_1] \nnb \\
\kmp m_{\Lambda_b} \, 
\mbox{\rm Re}[A_1^\ast E_3 - A_2^\ast E_1 + B_1^\ast D_3 -
B_2^\ast D_1] + m_{\Lambda_b}^2 (1-\hat{r}_\Lambda) \Big(                 
\vel A_2 \ver^2 - \vel B_2 \ver^2 \Big) \nnb \\
\ar m_{\Lambda_b} \sqrt{\hat{r}_\Lambda} \, \mbox{\rm Re}
[2 A_1^\ast A_2 -  2 B_1^\ast B_2 \mp A_1^\ast D_3 \mp 
A_2^\ast D_1 \mp B_1^\ast E_3 \mp B_2^\ast E_1] \nnb \\
\ek {(1-\hat{r}_\Lambda) \over \hat{s}}
\Big( \pm \mbox{\rm Re}[A_1^\ast D_1 + B_1^\ast E_1]
\Big)
\Bigg\}
~, \nnb \\
\nnb \\ \nnb \\
\label{e111225}
P_N^\mp \es {16 \pi m_{\Lambda_b}^3 \hat{m}_\ell v \sqrt{\hat{s}\lambda} \over \Delta} 
\Bigg\{ \pm \mbox{\rm Im}[A_1^\ast D_1 - B_1^\ast E_1]
\nnb \\
\ar m_{\Lambda_b} \Big(
\pm \mbox{\rm Im}[B_1^\ast D_2 - A_1^\ast E_2] + 
\mbox{\rm Im}[(\pm A_2+D_2+D_3)^\ast E_1] \nnb \\ 
\ek \mbox{\rm Im}[(\pm B_2-E_2-E_3)^\ast D_1] \Big) \nnb \\
\kmp m_{\Lambda_b} \Big(
m_{\Lambda_b} (1-\hat{r}_\Lambda) \, \mbox{\rm Im}[A_2^\ast D_2-B_2^\ast E_2] +
\sqrt{\hat{r}_\Lambda} \, \mbox{\rm Im}[A_1^\ast D_2 +A_2^\ast D_1] \Big) \nnb \\
\ar m_{\Lambda_b} \sqrt{\hat{r}_\Lambda} \,       
\mbox{\rm Im}[D_1^\ast (D_2-D_3) - E_2^\ast (\pm B_1+E_1) -
E_1^\ast (\pm B_2+E_3)]
\Bigg\}
~.\nnb
\eea

Using these definitions for the doubly--polarized $FB$ asymmetries, we get
the following results:   

\bea
\label{e111220}
{\cal A}_{FB}^{LL} \es \frac{32 m_{\Lambda_b}^5 \hat{s}\sqrt{\lambda} v}{\Delta}
\mbox{\rm Re} \Big[
- \Big\{  m_{\Lambda_b} (1-\hat{r}_\Lambda ) (A_2 D_2^\ast -
B_2 E_2^\ast) + \sqrt{\hat{r}_\Lambda} (A_1 D_2^\ast + A_2 D_1^\ast) 
\Big\} \nnb \\
\ar \sqrt{\hat{r}_\Lambda}
( B_1 E_2^\ast + B_2 E_1^\ast )
\Big]~, \\ \nnb \\
\label{e111221}
{\cal A}_{FB}^{LT} \es - {\cal A}_{FB}^{TL} = 
\frac{64m_{\Lambda_b}^4 \hat{m}_\ell \lambda }{3 \sqrt{\hat{s}}\Delta}
\mbox{\rm Re} \Big[
- \Big\{ \vel A_1 \ver^2 + \vel B_1 \ver^2 \Big\}
+ m_{\Lambda_b}^2 \hat{s}  
\Big\{ \vel A_2 \ver^2 + \vel B_2 \ver^2 \Big\}
\Big]~, \\ \nnb \\
\label{e111223}
{\cal A}_{FB}^{LN} \es {\cal A}_{FB}^{NL} = 
\frac{64 m_{\Lambda_b}^4 \hat{m}_\ell \lambda v }{3 \sqrt{\hat{s}}\Delta}
\mbox{\rm Im} \Big[
- ( A_1 D_1^\ast + B_1 E_1^\ast )
+ m_{\Lambda_b}^2 \hat{s}
( A_2 D_2^\ast + B_2 E_2^\ast )
\Big]~, \\ \nnb \\
\label{e111225}
{\cal A}_{FB}^{NT} \es {\cal A}_{FB}^{TN} =
\frac{64 m_{\Lambda_b}^4
\hat{m}_\ell^2 \sqrt{\lambda}}{\hat{s} \Delta}
\mbox{\rm Im} \Big[
m_{\Lambda_b} \hat{s} 
\Big\{ A_1 E_3^\ast - A_2 E_1^\ast + B_1 D_3^\ast 
- B_2 D_1^\ast \Big\} \nnb \\
\ar m_{\Lambda_b} \hat{s} \sqrt{\hat{r}_\Lambda}
( A_1 D_3^\ast + A_2 D_1^\ast + B_1 E_3^\ast +B_2 E_1^\ast ) 
+( 1 - \hat{r}_\Lambda)
( A_1 D_1^\ast + B_1 E_1^\ast ) \nnb \\
\ek m_{\Lambda_b}^2 \hat{s}^2
( A_2 D_3^\ast + B_2 E_3^\ast )
\Big]~, \\ \nnb \\
\label{e111226}
{\cal A}_{FB}^{NN} \es {\cal A}_{FB}^{TT}  = 0~.
\eea

In the expressions for ${\cal A}_{FB}^{ij}$, the superscript indices
$i$ and $j$ correspond to the lepton and anti--lepton polarizations, 
respectively, and $\Delta$ is determined from the differential decay
rate,
\bea
\label{nolabel}
{d\Gamma \over d\hat{s}} = {G_F \alpha_{em}^2 \over 8192 \pi^5} \vel
V_{tb}V_{ts}^\ast \ver^2 v \sqrt{\lambda(1,\hat{r}_\Lambda,\hat{s})}
\Delta~. \nnb
\eea
In all expressions the quantities $\lambda(1,\hat{r}_\Lambda,\hat{s})$,
$\hat{s}$,
$\hat{r}_\Lambda$, $\hat{m}_\ell$ and $v$ are defined as
$\lambda(1,\hat{r}_\Lambda,\hat{s}) = 1 + \hat{r}_\Lambda^2 + \hat{s} - 2 \hat{r}_\Lambda 
- 2 \hat{s} - 2 \hat{r}_\Lambda \hat{s}$, $\hat{s} = q^2/m_{\Lambda_b}^2$,
$\hat{r}_\Lambda = m_\Lambda/m_{\Lambda_b}$, $\hat{m}_\ell = m_\ell/m_{\Lambda_b}$,
and $v = \ds \sqrt{1 - {4 \hat{m}_\ell^2 \over \hat{s}}}$.

\section{Numerical analysis}

In the previous section we present the expressions for double and single
lepton polarizations in family non--universal $Z^\prime$ model. We now
proceed with the numerical analysis of these physical observables. In
addition to the input parameters in the SM, the considered version of the 
family non--universal $Z^\prime$ model contains four new parameters, namely,
$\vel B_{sb}^L \ver$, $\varphi_s^L$, $B_{\ell\ell}^L$ $B_{\ell\ell}^R$. The
constraints to these parameters coming from the analysis of present
experimental data in the $B$ meson sector are studied in detail in the
literature \cite{R111222}.

The values of the new input parameters appearing in family non--universal
$Z^\prime$ model are given in Table 1, in which S1 and S2 correspond to
UT--fit Collaboration result \cite{R111223}.

\begin{table}[h]    
\renewcommand{\arraystretch}{1.5}
\addtolength{\arraycolsep}{3pt}
$$
\begin{array}{|l|c|c|c|c|}
\hline
 & \vel B_{sb}^L \ver\times 10^{-3} & \varphi_s^{L[0]} & 
S_{\mu\mu}^L \times 10^{-2} & {\cal D}_{\mu\mu}^R\times 10^{-2} \\ \hline
S1 & 1.09 \pm 0.22 &  -72 \pm 7 & -2.80 \pm 3.90 & -6.70 \pm 2.60  \\ \hline
S2 & 2.20 \pm 0.15 &  -82 \pm 4 & -1.20 \pm 1.40 & -2.50 \pm 0.90  \\ \hline
\end{array}
$$
\caption{The values of four input parameters appearing in family
non--universal $Z^\prime$ model.}
\renewcommand{\arraystretch}{1}
\addtolength{\arraycolsep}{-3pt}
\end{table}

We have studied the sensitivities of single single and double lepton
polarizations on input parameters of family non--universal $Z^\prime$ model.
We can summarize the result of our analysis as follows:

\begin{itemize}

\item $P_L$ decreases maximally \%5 in both scenarios compared to the SM
prediction.

\item The values of $P_T$ and $P_N$ practically do not change. Therefore we
can conclude that single lepton polarization effects are not so efficient
for establishing new physics in the framework of family non--universal
$Z^\prime$ model.

\end{itemize}

As a result of the analysis of double lepton polarization we obtain that:

\begin{itemize} 

\item Predictions for $P_{LL}$, $P_{LT}$, $P_{TL}$ do coincide for both SM
and family non--universal $Z^\prime$ model.

\item Double lepton polarizations $P_{NN}$ and $P_{TT}$ are quite sensitive
to the parameters of $Z^\prime$ model. We present the $q^2$ dependence of 
$P_{NN}$ and $P_{TT}$ in Figs. (1) and (2), respectively. We observe from
these figures that, in the region $3~GeV^2 \le q^2 \le 15~GeV^2$ there
occurs considerable difference between the predictions of the SM and family
non--universal $Z^\prime$ model. Especially, the predictions of S1 scenario
for $P_{NN}$ and $P_{TT}$ shows larger discrepancy compared to S2.

\item In Fig. (3) we present the dependence of the polarized
forward--backward asymmetry ${\cal A}_{LL}$ on $q^2$ in the SM and family
non--universal $Z^\prime$ model. It follows from this figure that the zero
position of ${\cal A}_{LL}$ is shifted to left compared to the prediction of
the SM. Therefore determination of the zero position of ${\cal A}_{LL}$ can
give invaluable information, not only about the existence of new physics,
but also about the discrimination of the scenarios S1 and S2.

\end{itemize}   

We have also analysed the remaining forward--backward asymmetries 
${\cal A}_{FB}^{LN}$, ${\cal A}_{FB}^{NL}$, ${\cal A}_{FB}^{LT}$, ${\cal
A}_{FB}^{TL}$, ${\cal A}_{FB}^{NT}$ and ${\cal A}_{FB}^{TN}$ and obtained
that the contribution of new $Z^\prime$ bosons to these asymmetries are
negligibly small.

As the concluding remark we can summarize our analysis as follows.
Contributions of family non--universal $Z^\prime$ model to the single and
double lepton polarizations, as well as polarized forward--backward
asymmetry ${\cal A}_{LL}$ in rare, exclusive baryonic $\Lambda_b \to \Lambda
\ell^+ \ell^-$ decay is studied. It is obtain that $P_{NN}$ and $P_{TT}$
are quite sensitive to the $Z^\prime$ boson contributions. Moreover, it is
found that zero position of the forward--backward asymmetry ${\cal A}_{LL}$
is shifted to left compared to the SM case. Determination of the value of
zero position of ${\cal A}_{LL}$ is also a very important information for
the scenarios under consideration. The results we obtain can all be checked
in future planned LHC--b experiments.

\newpage

\section*{Figure captions}
{\bf Fig. (1)} The dependence of the double--lepton polarization asymmetry
$P_{LL}$ on $q^2$ for the $\Lambda_b \rar \Lambda \mu^+ \mu^-$ 
decay. \\ \\
{\bf Fig. (2)} The same as in Fig. (1), but for the double--lepton
polarization asymmetry $P_{TT}$.\\ \\
{\bf Fig. (3)}  The dependence of the double--lepton polarization asymmetry
${\cal A}_{FB}^{LL}$ on $q^2$ for the $\Lambda_b \rar \Lambda \mu^+ \mu^-$ 
decay.

\newpage

\begin{figure}
\vskip 1.5 cm
    \includegraphics{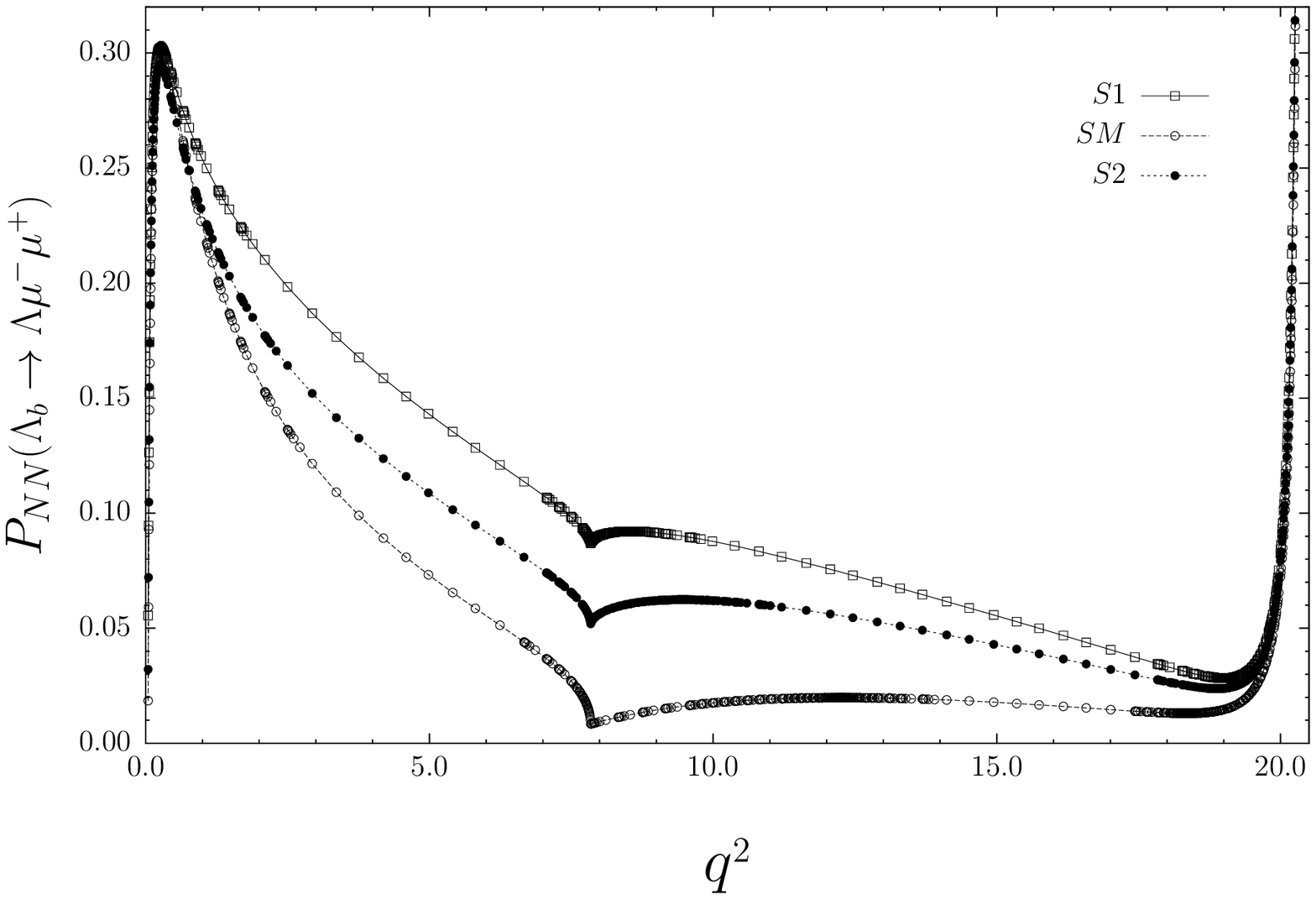}
\vskip 7.8cm
\caption{}
\end{figure}  

\begin{figure}   
\vskip 2.5 cm
    \includegraphics{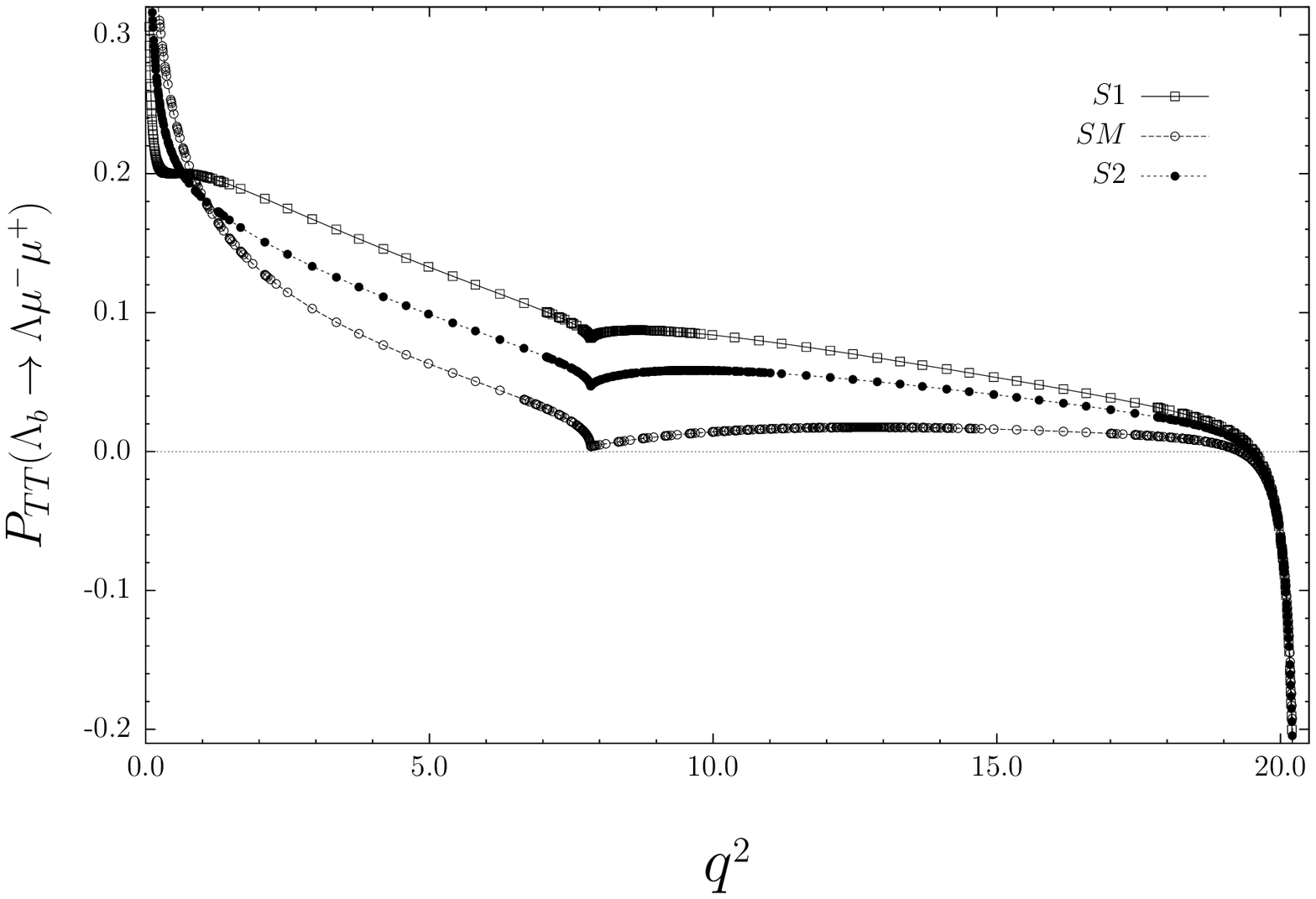}
\vskip 7.8 cm   
\caption{}
\end{figure}

\begin{figure}   
\vskip 1.5 cm
    \includegraphics{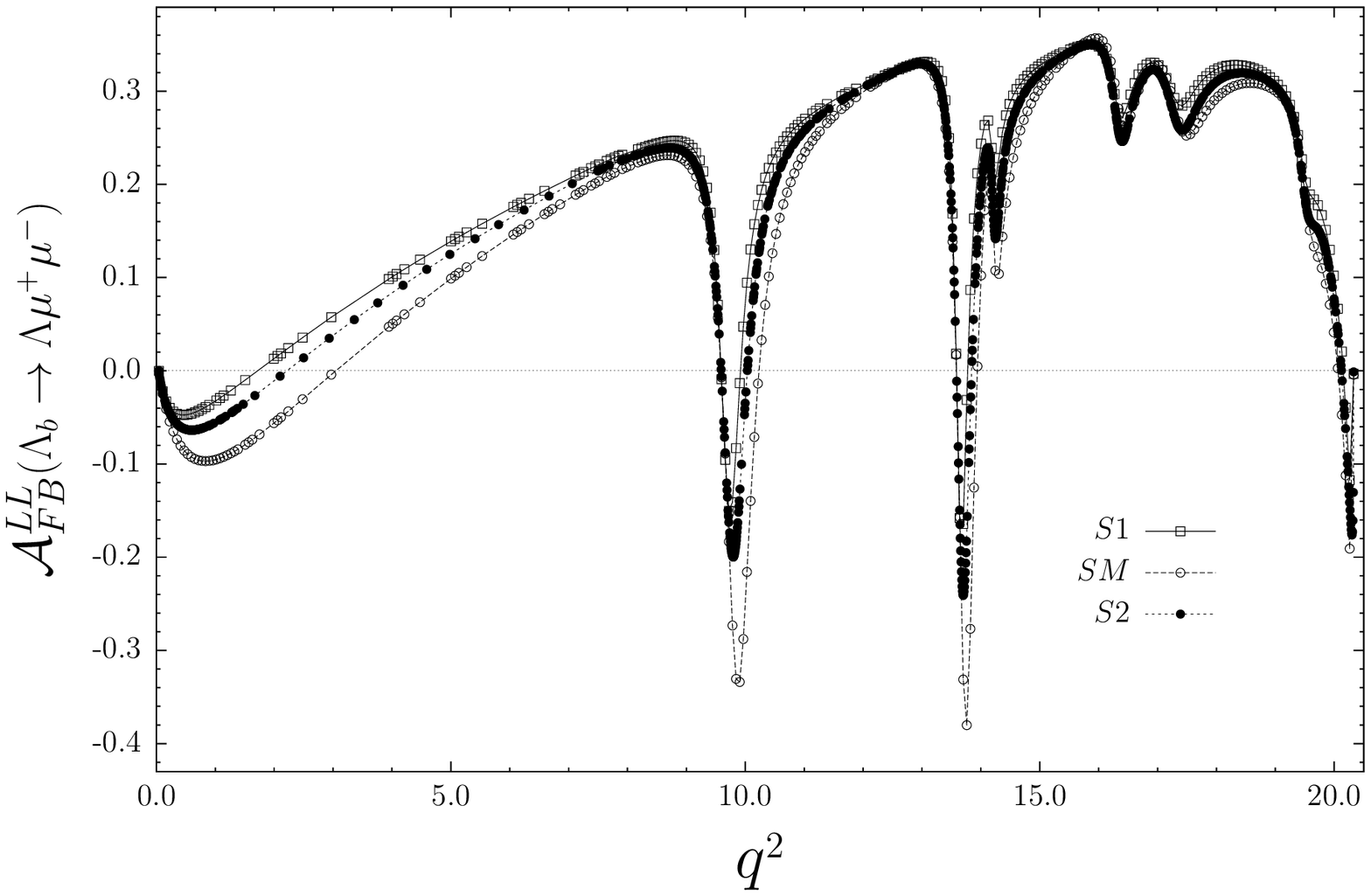}
\vskip 7.8cm
\caption{}
\end{figure}

\end{document}